\documentclass[preprint,12pt]{elsarticle}
\usepackage{tabularx}
\usepackage{amssymb}
\usepackage{graphics}
\usepackage[dvips]{epsfig}
\usepackage{fancyhdr}
\usepackage{threeparttable}

\author[a]{Xiangde Zhu\corref{cor1}}
\author[b]{Wenjian Lu\corref{cor1}}
\author[a]{Wei Ning}
\author[a]{Zhe Qu}
\author[c]{Li Li}
\author[c]{T. F. Qi}
\author[c]{Gang Cao}
\author[d]{Cedomir Petrovic}
\author[a]{Yuheng Zhang}
\address[a]{High Magnetic Field Laboratory, Chinese Academy of Sciences, Hefei 230031, China}
\address[b]{Key Laboratory of Materials Physics, Institute of Solid State Physics, Chinese Academy of Sciences, Hefei 230031, China}
\address[c]{Department of Physics and Astronomy and Center for Advanced Materials, University of Kentucky, Lexington, KY 40506, USA}
\address[d]{Condensed Matter Physics and Materials Science Department, Brookhaven National Laboratory, Upton, NY 11973, USA}
\cortext[cor1]{Corresponding authors at: High Magnetic Field Laboratory, Chinese Academy of Sciences, Hefei 230031, China. TEl: +86-551-65595619; FAX: +86-551-65591149.
Email: xdzhu@hmfl.ac.cn(Experiment);wjlu@issp.ac.cn(Calculation)}

\title
{Single Crystal Growth, Transport, and Electronic Band Structure of YCoGa$_5$}

\begin{document}
\begin{frontmatter}
%%%%%%%%%%%%%%%%%%%%%%%%%%%%%%%%%%%%%%%%%%%%%%%%%%%%%%%%%%%%%%%%%%%%%
\begin{abstract}
Single crystal of YCoGa$_5$ has been grown via Ga self-flux. In this paper, we report the single crystal growth, crystallographic parameters, resistivity, heat capacity, and band structure results of YCoGa$_5$. YCoGa$_5$ accommodates the HoCoGa$_5$ type structure (space group P4/mmm (No. 123), Z = 1, a = 4.2131(6) {\AA}, c = 6.7929(13) {\AA}, which is isostructural to the extensively studied heavy fermion superconductor system Ce$M$In$_5$ ($M$ = Co, Rh, Ir) and the unconventional superconductor PuCoGa$_5$ with $T_c$ = 18.5 K. No superconductivity is observed down to 1.75 K. Band structure calculation results show that its band at the Fermi level is mainly composed of Co-3d and Ga-4p electrons states, which explains its similarity of physical properties to YbCoGa$_5$ and LuCoGa$_5$.

\end{abstract}


\end{frontmatter}
\section{Introduction}
New materials may stimulate extensive research, and could clarify or bring forth important questions about the underlying physics. For example, CeCoIn$_5$, the heavy fermion superconductor with the highest superconducting transition temperature ($T_c$) of 2.3 K, was discovered by Petrovic $et~al.$ in 2001. It became the most extensively studied heavy fermion system during the past decade.\cite{CeCoIn5} Later in 2002, Sarrao $et~al.$ reported the discovery of 5$f$ superconductor PuCoGa$_5$ with $T_c$ =18.5 K.\cite{PuCoGa5} These two compounds strongly support the unconventional superconductivity mediated by magnetic fluctuations.\cite{USC} Both of them belong to the ternary HoCoGa$_5$-type tetragonal structure (space group P4/mmm), which can be regarded as alternative stacking of HoGa$_3$-CoGa$_2$-HoGa$_3$ layers sequentially along the $c$-axis. This leads to rather unique and quasi two dimensional electronic states. Due to the heavy fermion superconductor CeMIn$_5$ (M is Co, Rh, Ir) materials and the wide family of $Ln$TX$_5$ and $An$TX$_5$ (here, $Ln$ is the rare earth atoms, $An$ is the actinide, T is transition metal, X is In or Ga),\cite{CeIrIn5,CeRhIn5,RCoGa5,RMIn5,RMGa5} the HoCoGa$_5$ type structure have attracted attention both from physics and chemistry fields. So far, only Gd, Tb, Dy, Ho, Er, Tm, Yb, Lu and Y can form HoCoGa$_5$ type strucutre.\cite{RMGa5} The single crystals for Tb-Tm were grown by Hudis $et~al$, showing metallic behavior with Kondo antiferromagnetism.\cite{RCoGa5} The single crystal for Yb(4f$^{14}$6S$^2$)CoGa$_5$ and Lu(4f$^{14}$5d$^1$6S$^2$)CoGa$_5$ were grown by Okudzeto $et~al.$ \cite{YbCoGa5} and Matsuda $et~al.$\cite{LuCoGa5}, respectively. Although Yb usually shows valence fluctuations, spin fluctuations and quantum critical behavior in many Yb-based intermetallic compound,\cite{VLF,VLP,YbRh2Si2,YbCuGa,YbCuIn,YbAl3,YbRhSn} YbCoGa$_5$ shows typical simple metallic behavior similar to LuCoGa$_5$.\cite{YbCoGa5,LuCoGa5} Thus, Yttrium without f electrons may provide a good reference to study structure-property relations in 115 Ga compounds. So far, we have noticed that YCoGa$_5$ has no report on single crystal or detailed physical properties in our knowledge, except for the structure report.\cite{RMGa5}.

In addition, 13 phase compound CeIn$_3$ (space group $Pm\overline{3}m$, a = 4.689 {\AA}) has also attracted some attentions due to its tunable superconductivity and antiferromagnetism under pressure.\cite{CeIn3} The iso-structural ScGa$_3$ and LuGa$_3$ are both type I superconductors.\cite{ScGa3} Unfortunately, YGa$_3$ does not exist. Thus, it is interesting to search for the superconductivity in the YCoGa$_5$, since its structure is composed of YGa$_3$-CoGa$_2$-YGa$_3$ layers. In this paper, we report single crystal growth, crystallographic parameters, physical properties and electronic structure of YCoGa$_5$.

\section{Material and methods}

  \textbf{Single Crystal Growth.} The single crystal of YCoGa$_5$ was grown via gallium self-flux method. Yttrium shot (Alfa Aesar, 99.9\%), Cobalt powder (Aladin, 99.9\%), and Gallium shot (99.995\%) in 1:1:30 mol ratio were put in an 2 mL aluminum crucible, and sealed in an evacuated quartz tube. The tube was heated to 1100 $^\circ$C, and dwelled for 6 h, then cooled down to 400 $^\circ$C in a 2 K/h rate. At this temperature, crystals were separated from the flux using a centrifuge. Single crystal with shiny face can be found with the size of 5 $\times$ 5 $\times$ 3 mm$^3$, which is shown in Fig. 2. In order to remove the Ga flux on the surface, the sample was put in boiling water for about 30 minutes, and then the etching in dilute HCl for several hours.

\begin{figure}[htbp]
  \centering
  \includegraphics[width=0.45\textwidth,angle=0]{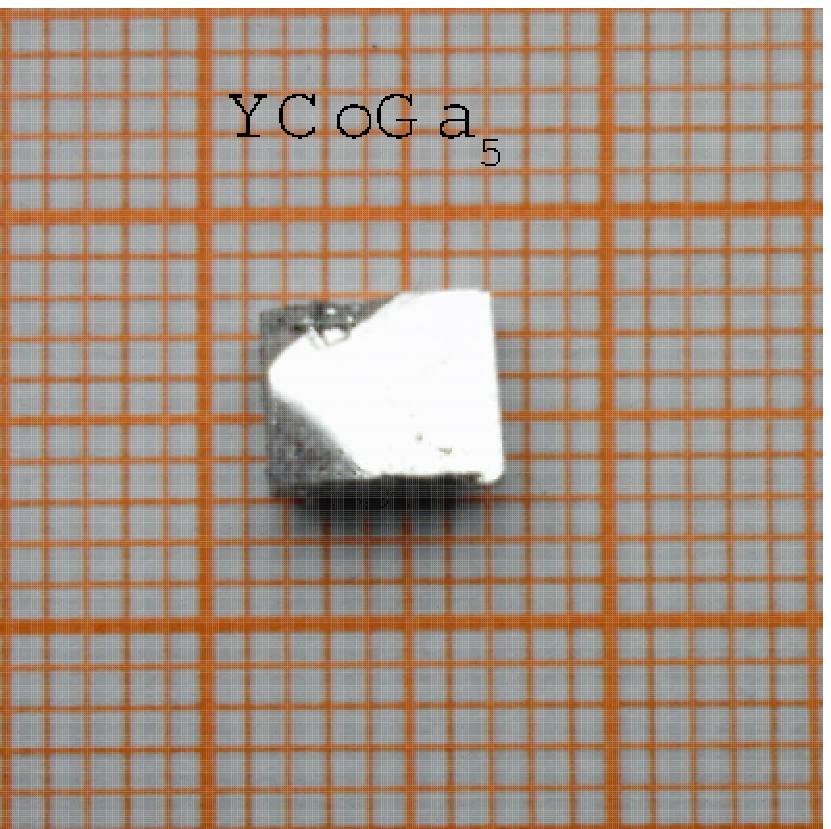}\\
  \caption{The photograph of as-grown YCoGa$_5$ single crystal.}
\end{figure}

  \textbf{Single X-ray Diffraction and Elemental Analysis.}
  Crystal fragments with dimensions of about 0.1 $\times$ 0.1 $\times$ 0.1 mm$^3$ were mechanically cut from the crystal and selected for the single crystal X-ray experiments of YCoGa$_5$. The crystal was glued on the tip of a quartz fiber and mounted on a Nonius Kappa CCD diffractometer outfitted with a Mo Kappa radiation ($\lambda$ ) = 0.71073 {\AA}). Data collection was taken at 296 K. The crystal structure of YCoGa$_5$ was solved using SHELXS97 and refined using SHELXL97.\cite{Shelx} The crystallographic parameters are listed in Table 1. After refinement the data were corrected for extinction effects and the displacement parameters were refined as anisotropic. A list of atomic positions, Wyckoff symmetry, and anisotropic displacement parameters are shown in Table 1. The composition of the single crystal was determined by energy dispersive X-ray spectrum (EDS) on Hitachi TM3000 Tabletop Microscope system. The average elemental ratio is Yb/Co/Ga $\sim$ 1.1:1:4.93.

  \textbf{Transport Properties.} Electrical transport measurement was carried out on a Linear Research AC bridge and a Quantum Design Magnetism Property Measurement System, with the current applied in the $ab$-plane of the sample. Data were collected over a temperature range of 1.75 - 300 K. Temperature dependent resistivity measurement was performed using a four-probe configuration. Gold wires were attached to a polished bar-shaped sample with the electrical contacts made by silver epoxy. Heat capacity measurement was carried out on a Quantum Design Physical Property Measurement System PPMS-16.

  \textbf{Band Calculations.} The electronic strucutre of YCoGa$_5$ were calculated by first principles within density functional theory (DFT). The calculations were performed by means of the projected augmented-wave (PAW)\cite{blochl1994,torrent2008} formalism implemented in the ABINIT code.\cite{gonze2009,gonze2002,gonze_brief_2005} The exchange-correlation functions are treated by using the generalized gradient approximation (GGA) according to the Perdew-Burke-Ernzerhof (PBE)\cite{perdew1996} parametrization. The basis set of valence electronic states was taken to be $4s^24p^64d^15s^2$ for Y, $3s^23p^63d^74s^2$ for Co, and $3s^23p^63d^{10}4s^24p^1$ for Ga, respectively. Electronic wave functions are expanded with plane waves up to an energy cutoff ($E_{cut}$) of 1400 eV. Brillouin zone sampling is performed on a Monkhorst-Pack (MP) mesh\cite{monkhorst1976} of $16\times16\times12.$ The self-consistent calculations were considered to be converged when the total energy of the system was stable within $10^{-6}$ eV per atom.

\section{Results and Discussions}

\subsection{Crystal Structure}

As shown in Fig. 2(a), the crystal structure of YCoGa$_5$ is composed of alternative YGa$_3$ cuboctahedra layers and CoGa$_2$ rectangular prisms layers stacking along $c$-axis. Figure 2 (b), (c), (d) and (e) shows (0kl), (h0l), (hk0) and (hk3) reciprocal planes from the single crystal X-ray diffraction patterns at $T$ = 296 K, respectively. Obviously, the well-defined patterns indicate that the single crystal is of high quality. The crystallographic and atom parameters for YCoGa$_5$ obtained from Rietveld Refinement are listed in Tab. 1. The lattice parameters are consistent with the previous report. All the parameters for YCoGa$_5$ are very close to those for YbCoGa$_5$ and LuCoGa$_5$.\cite{YbCoGa5,LuCoGa5}

\begin{figure}[htbp]
  \centering
  \includegraphics[width=0.5\textwidth,angle=0]{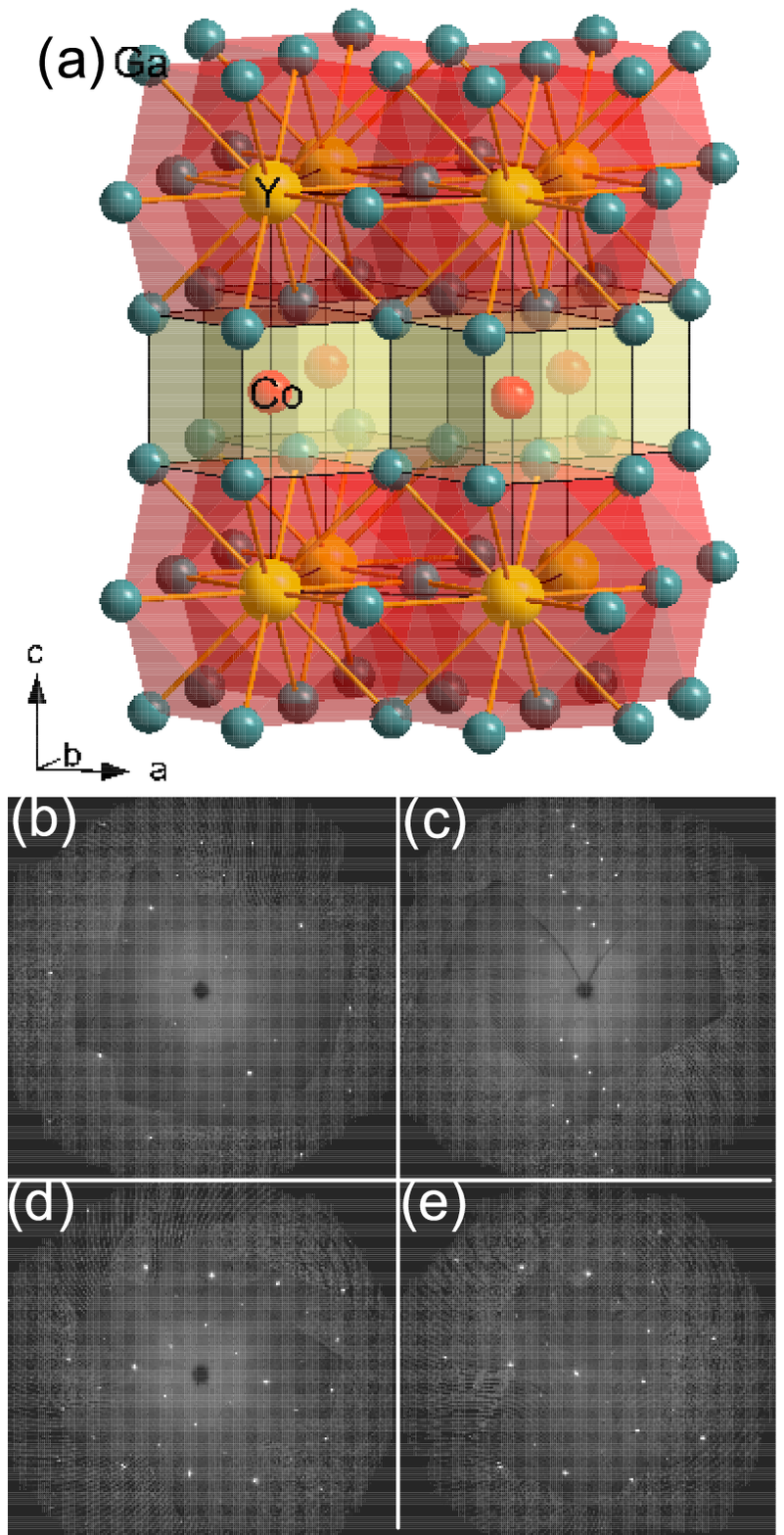}\\
  \caption{(a) The crystal structure of YCoGa$_5$. (b), (c), (d), (e) shows (0kl), (h0l), (hk0), (hk3) reciprocal planes from the single crystal X-ray diffraction pattern at T = 296 K, respectively. }
\end{figure}

\begin{table}[htb]\centering%
\caption{Crystallographic parameters for YCoGa$_5$.} \label{Tab:1}
\label{tbl:1}
\begin{threeparttable}
 \begin{tabular}{cc}
 \hline
  Parameters & YCoGa$_5$  \\
  \hline
  S.G. & P4/$mmm$(123)   \\
  a & 4.2131(6) {\AA}    \\
  c & 6.7929(14) {\AA}   \\
  c/a& 1.612             \\
  V( {\AA}$^3$)& 120.58(3)\\
  Z & 1    \\
  dimensions (mm)& 0.1$\times$0.1$\times$0.1\\
  Density & 6.837 g cm$^{-3}$ \\
  Temperature &  296 K  \\
  $\theta$ range & 3-27.4$^\circ$ \\
  collected reflections & 114 \\
  $h$ & -3 $\leq h \leq$ 3  \\
  $k$ & -5 $\leq k \leq$ 5  \\
  $l$ & -8 $\leq l \leq$ 8  \\
  $\Delta\rho_{max}$ (e {\AA}$^{-3}$)& 2.70\\
  $\Delta\rho_{min}$ (e {\AA}$^{-3}$)& -1.54\\
  $^aR_1$[F$^2 > 2\sigma$(F$^2$)] & 0.035 \\
  $^b$$R_w$ & 0.109         \\
 \end{tabular}
      \begin{tablenotes}
        \footnotesize
        \item[a]{$R_1$ = $\sum||F_o| - |F_c||/\sum|F_0|$. }
        \item[b]{$R_w = [\sum[w(F_o^2 - F_c^2)2]/\sum[w(F_o^2)^2]]^{1/2}$; $w = 1/[\sigma^2(F_o^2) + (0.065P)^2 + 0.5173P]$; $P$ = ($F_o^2 + 2F_c^2$)/3.}
      \end{tablenotes}
     \end{threeparttable}
\end{table}

\begin{table}
  \caption{The fractional coordinates and thermal parameters for YbCoGa$_5$. Here, $U_{eq}$ is defined as $1/3$ of the trace of the orthogonalized $U_{ij}$ tensor.}\label{Tab:2}
 \begin{tabular}{cccccc}
  \hline
atom & Wyckoff & x & y &z & $U_{eq}$\\
 \hline
Y   & 1a  &   0& 0 & 0 & 0.0011(7) \\
Co  & 1b  &   0& 0 &1/2& 0.0029(7)\\
Ga1 & 1c  & 1/2&1/2& 0 & 0.0112(7)\\
Ga2 & 4i  &   0&1/2&0.3107(2)& 0.0118(6) \\
  \end{tabular}\\
\end{table}

\subsection{Transport Properties}

\begin{figure}[tbp]
  \centering
  \includegraphics[width=0.5\textwidth]{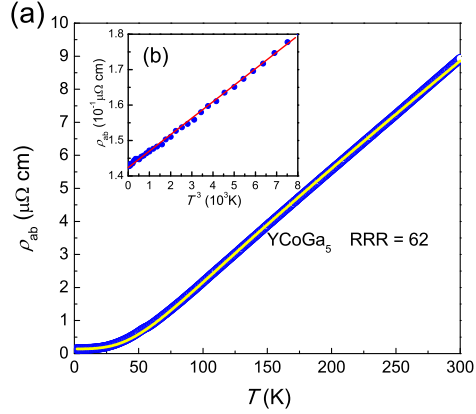}\\
  \caption{(a) The temperature dependence of in-plane resistivity ($\rho_{ab}$, circle) and its fitting results (solid line) for YCoGa$_5$. (b) The $\rho_{ab}$ versus $T^3$ at low temperature for YCoGa$_5$. The solid line serves as a guide of eye.}
\end{figure}

Figure 3(a) shows the temperature dependence of in-plane resistivity ($\rho_{ab}$) for YCoGa$_5$. Obviously, it shows a typical metallic behavior, with a small residual resistivity ($\rho_0$) of 0.14 $\mu \Omega$ cm. The obtained residual resistivity ratio (RRR) is $\sim$ 62, indicating the high quality of the single crystal. These values are comparable with those of YbCoGa$_5$ and LuCoGa$_5$.\cite{YbCoGa5,LuCoGa5} As shown in Fig. 3(b), the linear relationship of $\rho$($T$)-$T^3$ can be observed, rather than $\rho$($T$)-$T^5$ for usual electron-phonon scattering.  This demonstrates that the $s-d$ electron scattering is the major scattering mechanism in YCoGa$_5$. As shown in Fig. 3(a), $\rho_{ab} - T$ curve can be fitted with the Bloch-Gr$\ddot{u}$neisen formula with $n~\sim$ 3
\begin{equation}
\rho(T) = \rho_0+2A_{ac}(\frac{T}{\Theta_R})^{n-1}T\int^{\frac{\Theta_R}{T}}_{0}\frac{x^n}{(e^x-1)(1-e^{-x})}dx
\end{equation}
 in the whole range, where $\Theta_R$ is the Debye temperature obtained from the resistivity measurements, $A_{ac}$ is the temperature dependent coefficient. The fitting gives $A_{ac}$ = 0.061223 $\mu\Omega$ cm K$^{-1}$, and $\Theta_R$ = 338 K. This type of $\rho -T$ relation has been observed in LuIn$_3$,\cite{LuIn3} and LuGa$_3$.\cite{ScGa3}

\begin{figure}[tbp]
  \centering
  \includegraphics[width=0.5\textwidth]{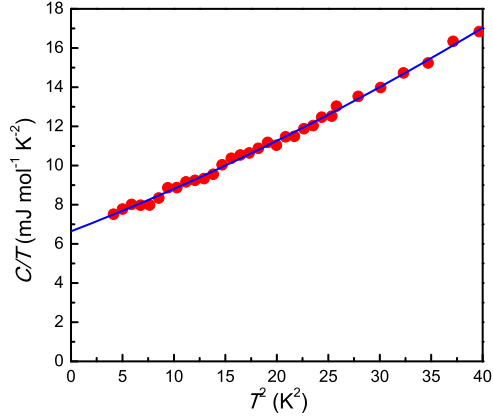}\\
  \caption{The $T^2$ dependence of specific heat ($C$) divided by $T$ for YCoGa$_5$. The solid line shows the fitting results.}
\end{figure}

Figure 4 depicts the $T^2$ dependence of specific heat($C$) divided by $T$ for YCoGa$_5$. The data can be well fitted by the formula $C = \gamma{T}+\beta{T^3}+\delta{T^5}$, where $\gamma{T}$ is the electron contribution, $\beta{T^3}$ is the phonon contribution, and $\delta{T^5}$ is the anharmonic contribution. The obtained results from the fitting gives $\gamma$ = 6.64 $\pm$ 0.11 mJ mol$^{-1}$ K$^{-2}$, $\beta$ = 0.202 $\pm$ 0.012 mJ mol$^{-1}$ K$^{-4}$, and $\delta$ = 1.4 $\pm$ 0.3 mJ mol$^{-1}$ K$^{-6}$. $\Theta_D$ = 407 $\pm$ 8K is estimated from the $\Theta_D = \sqrt[3]{12\pi^4{N}R/5\beta}$, where N is the number of atoms per molecule.

All the lattice and physical parameters for single crystal YCoGa$_5$, YbCoGa$_5$, LuCoGa$_5$ and PuCoGa$_5$ are listed in Tab. 3. The lattice parameters and the $c/a$ ratios are very close to each other, as well as the physical parameters except PuCoGa$_5$. Therefore, it is worthwhile to figure out the physics origin of the similarity between YCoGa$_5$, YbCoGa$_5$ and LuCoGa$_5$. Band structure calculation will provide good opportunity to understand their physical properties.

\begin{table}[htbp]\centering
\caption{The lattice and physical parameters for single crystal YCoGa$_5$, YbCoGa$_5$, LuCoGa$_5$ and PuCoGa$_5$. }
\label{Tab:3}
\begin{threeparttable}
\begin{tabular}{ccccc}
  \hline
                                  & YCoGa$_5$  & YbCoGa$_5$ & LuCoGa$_5$ & PuCoGa$_5$ \\
  \hline \\
  a(\AA)                          & 4.2131(6)      & 4.190(1)     & 4.184(3)      & 4.232      \\
  c(\AA)                          & 6.7929(14)     & 6.727(3)     & 6.718(7)      & 6.786      \\
  c/a                             & 1.6122(6)      & 1.605(6)     & 1.606(3)      & 1.603      \\
  $\rho_0$($\mu\Omega$ cm)        & 0.14       & 0.13       & 0.31       & $\sim$ 35  \\
  RRR                             &  62        & 150        & 52         & $\sim$ 7   \\
  $\gamma$ (mJ mol$^{-1}$ K$^{-2}$)& 6.64      & 11.2       & $\sim 6$   & 77         \\
  $\Theta_D$(K)                   &  408       & 420$^a$    & 344$^a$    &            \\
  Referece                        &  This work & Ref\cite{YbCoGa5}& Ref\cite{LuCoGa5} &Ref\cite{PuCoGa5}\\
\end{tabular}
      \begin{tablenotes}
        \footnotesize
        \item[a] The $\Theta_D$ obtained in Ref\cite{YbCoGa5} and Ref\cite{LuCoGa5} are both calculated using formula $\Theta_D = \sqrt[3]{12\pi^4R/5\beta}$. Here shows the corrected values.
      \end{tablenotes}
\end{threeparttable}
\end{table}

\subsection{Band Calculations}
\begin{figure}[tbp]
  \centering
  \includegraphics[width = 0.5\textwidth ]{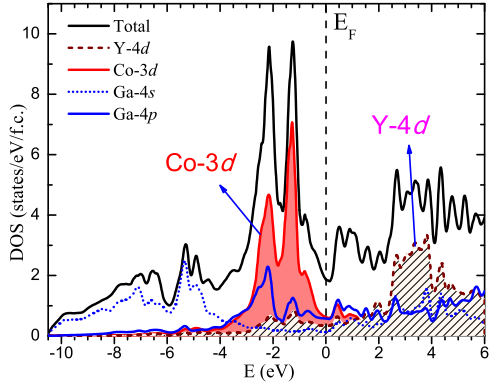}\\
  \caption{The calculated total and projected density of states (DOS) of YCoGa$_5$, where the partial density of states of Y-4d, and Co-3d are also shown as filled color.}\label{dos}
\end{figure}

The theoretically optimized lattice parameters of YCoGa$_5$ are established to be $a=4.227$~\AA~and $c=6.805$~\AA, which are very close to the experimental values. The band dispersion is shown in the supporting information. The calculated total and projected density of states (DOS) are shown in Fig.~\ref{dos}. The DOS in the region from -10 eV to -4 eV mainly consist of Ga-$4s$ states, which are far from the Fermi level $E_F$. There are two peaks below $E_F$ due to Co-$3d$ states that strongly hybridize with Ga-$4p$ states between -4 eV and 3 eV in valence and conduction bands. Due to the hybridization, the bandwidth of Co-$3d$ states becomes wide and results in an itinerant metallic behavior for YCoGa$_5$. As shown in Fig. 5, the DOS at Fermi level $N(E_F)$ mainly consist of Co-$3d$ and Ga-$4p$ states, while Y-$4d$ state palys a minor role; and the calculated value of $N(E_F)$ is 1.86 states/eV. In addition, the $E_F$ nearly locates at a local minimum in the total DOS, suggesting a higher structure stability of YCoGa$_5$, since it signifies a barrier for electrons below $E_F$ to move into unoccupied empty states.

As is shown in Tab. 3, YCoGa$_5$, YbCoGa$_5$ and LuCoGa$_5$ have similar lattice parameters. All of them show normal metallic behavior with small residual resistivity, non-magnetism, and non-superconductivity. For LuCoGa$_5$, the DOS at Fermi level $N(E_F)$ mainly consist of Co-$3d$ and Ga-$4p$ states, jus as the same as YCoGa$_5$.\cite{LuCoGa5} This strongly suggests that Yb should be divalent in YbCoGa$_5$ ($f$ electrons are below the fermi level), which explains the non-heavy-fermion and non-valence fluctuation in YbCoGa$_5$. Generally, the DOS at Fermi level $N(E_F)$ mainly consisting of Co-$3d$ and Ga-$4p$ states is the origin of the similarity between YCoGa$_5$, YbCoGa$_5$ and LuCoGa$_5$. In contrast, Pu 5$f$ electrons play an important role in PuCoGa$_5$, which confirms that the f-electrons and its magnetism is related to the superconductivity in PuCoGa$_5$.

\section{Conclusions}
High quality single crystals of YCoGa$_5$ were obtained from the Ga self-flux method. The lattice parameters, fractional coordinates, and thermal parameters of YCoGa$_5$ were determined from single-crystal X-ray diffraction measurements. The electrical resistivity and specific heat were measured, indicating that YCoGa$_5$ is a metal with similar physical properties to YbCoGa$_5$ and LuCoGa$_5$. The band structure calculation results demonstrate that its Fermi level mainly consist of Co-$3d$ and Ga-$4p$ states, which is as same as YbCoGa$_5$ and LuCoGa$_5$.

\section{Acknowledgements}
We are very grateful to Dr. Lei Zhang for the help on Powder X-ray experiments. Work at High magnetic field lab (Hefei) was supported by the State Key Project of Fundamental Research, China (2010CB923403), National Basic Research Program of China (973 Program), No. 2011CBA00111, and National Natural Science Foundation of China (Grants No. 11204312). Work at Brookhaven is supported by the US DOE under Contract No. DE-AC02-98CH10886.

\section{Supplementary data}
 Crystallographic information file for YCoGa$_5$.

 Figure: Band dispersions in YCoGa$_5$ along high-symmetry directions. The three bands cross Fermi level are shown in color lines.

\end{document}